\documentclass[11pt,preprint,superscriptaddress,aps,showkeys,nofootinbib]{revtex4}

\usepackage{amssymb,amsmath,amsfonts}
\usepackage{graphicx}
\usepackage{graphics}
\usepackage{eepic,epsfig}
\usepackage{indentfirst}
\usepackage[utf8]{inputenc}
\usepackage[T1]{fontenc}
\usepackage{subfig}

1.60cm \makeatletter \@addtoreset{equation}{section}

\makeatother

\begin{document}

\title{Polar gravitational perturbations and quasinormal modes of a loop quantum gravity black hole} 

\author{M. B. Cruz}
\email{messiasdebritocruz@gmail.com}
\affiliation{Departamento de F\'{\i}sica, Universidade Federal de Campina Grande \\
Caixa Postal 10071, 58429-900 Campina Grande, Para\'{\i}ba, Brazil}
\author{F. A. Brito}
\email{fabrito@df.ufcg.edu.br}
\affiliation{Departamento de F\'{\i}sica, Universidade Federal de Campina Grande \\
Caixa Postal 10071, 58429-900 Campina Grande, Para\'{\i}ba, Brazil}
\affiliation{Departamento de F\'{\i}sica, Universidade Federal da Para\'{\i}ba \\ Caixa Postal 5008, 58051-970, Jo\~ao Pessoa, Para\'{\i}ba, Brazil}
\author{C. A. S. Silva}
\email{carlosalex.phys@gmail.com}
\affiliation{Instituto Federal de Educa\c{c}\~{a}o Ci\^{e}ncia e Tecnologia do Cear\'{a} (IFCE),\\ Campus Tiangu\'{a} - Av. Tabeli\~{a}o Luiz Nogueira de Lima, s/n - Santo Ant\^{o}nio, Tiangu\'{a} - CE}
 
\begin{abstract}
In this work, we have calculated the polar gravitational quasinormal modes for a quantum corrected black hole model, that arises in the context of Loop Quantum Gravity, known as Self-Dual Black Hole. In this way, we have calculated the characteristic frequencies using the WKB approach, where we can verify a strong dependence with the Loop Quantum Gravity parameters. At the same time we check that the Self-Dual Black Hole is stable under polar gravitational perturbations, we can also verify that the spectrum of the polar quasinormal modes differs from the axial one \cite{Cruz:2015bcj}. Such a result tells us that isospectrality is broken in the context of Self Dual Black Holes.
\end{abstract}


\maketitle

\section{Introduction}

One of the most exceptional predictions of General Relativity (GR) \cite{Einstein:1915ca} is the existence of the black holes (BHs), that are objects from which nothing (even light signals) can escape after crossing their event horizons. The interest in BHs goes far beyond astrophysics. They have appeared as possible objects that can help us to understand one of the most intriguing problems in theoretical physics today: the nature of quantum gravity. This is because in the presence of a very strong gravitational field, the quantum properties of spacetime must become relevant.

Interest in BHs physics research has been boosted in the last years because of gravitational wave (GW) observations since 2015 \cite{Abbott:2016blz}, which was originated from a binary BHs merging, and since then this class of events has been observed with increasing of precision through the LIGO and Virgo collaborations \cite{TheLIGOScientific:2016htt, Abbott:2016nmj, Abbott:2017vtc, LIGOScientific:2020stg}. 

In the context of Loop Quantum Gravity (LQG), one of the main candidates to a theory of quantum gravity up to now, it is possible to find interesting theoretical models that provide an insight into the quantum characteristics of spacetimes revealed by BHs. Actually, in the context of this theory, an interesting BH scenario corresponds to the Self-Dual Black Hole (SDBH) \cite{Modesto:2009ve}, a quantum version of the Schwarzschild Black Hole (SchBH) that has the interesting property of self-duality. From such duality, the BH singularity is removed and replaced by an asymptotically flat region, which is an expected effect in the quantum gravity regime.

A very important fact is that we cannot find a BH completely isolated in nature, i.e., they are always interacting with other structures in its neighborhood (for instance, neutron stars and other black holes). However, even if we suppose that nothing exists near the BHs, they still will interact with the surrounding vacuum, creating pairs of particles, and consequently evaporating due to Hawking phenomena \cite{Hawking:1974sw}. Therefore, real BHs are always in a perturbed state and emitting GWs as a response to those perturbations. Such GWs are characterized by a set of complex eigenvalues of the wave equations, known as quasinormal modes (QNMs). In this context, the real part of the QNMs gives us the oscillation frequency, while the imaginary part determines the damping rates, and they depend only on BHs parameters (for instance, the mass and charge) and not on the sources that cause the perturbations.

The studies of the QNMs are of great interest and importance in different contexts. For example, in the context of the AdS/CFT, they are studied because of the possibility of observing the quasinormal ringing of astrophysical BHs, when one considers the thermodynamic properties of BHs \cite{Horowitz:1999jd}. Moreover, the QNMs of near extremal black branes have also been investigated in \cite{Berti:2009kk}. In recent years, it has been suggested that the QNMs might play an important role in quantum gravity theories, mainly in approaches like String Theory and Loop Quantum Gravity. Especially in the context of LQG, it has been also suggested that the QNMs can be used to fix the Immirzi parameter, a parameter measuring the quantum of the spacetime \cite{Dreyer:2002vy}. Also, this is a fundamental issue that remains open in this field. Inspired by these considerations, the QNMs of SDBHs have been computed in \cite{Santos:2015gja}, where scalar perturbations have been considered, and also in \cite{Cruz:2015bcj} where the present authors considered the axial gravitational spectrum.

In the context of quasinormal modes, an additional interesting issue is the question of isospectrality. In this way, some recent works have stressed such issues for black hole solutions beyond Schwarzschild \cite{Cardoso:2019mqo, Moulin:2019bfh}. Isospectrality consists of the coincidence between black hole axial and polar spectrum of quasinormal modes, and
it is a well-established property
for black holes in classical general relativity. However, it has been demonstrated that to assume isospectrality for black holes in the case of modified gravitational theories, like $f(R)$ or Chern-Simons gravity, consists of a naive stance. Results obtained in the context of such theories have shown that such property may not hold in this case.

The present work has as its goal to perform an analisys of the black hole quasinormal spectrum in the context of LQG. To do this, we shall examine the QNMs of an SDBH \cite{Modesto:2009ve}, by taking a model of a GW obtained from polar gravitational perturbations. To compute the QNMs spectrum we shall use the WKB approach, and compare them with literature results for axial perturbations \cite{Cruz:2015bcj}. In this context, we shall also analyze the stability of the self-dual solution over the polar gravitational perturbations, and analyze the implications of the quantum gravity corrections to isospectrality. As we shall see, isospectrality will be broken by SDBHs.

The paper is organized as follows. In Section \ref{intr_sel_dual_bh}, we briefly review the SDBH solution and discuss their self-duality property. In Section \ref{effect_potent}, we derive a Schrodinger-type equation, by considering the polar gravitational perturbations. In Section \ref{qnm_sdbh}, we shall calculate the QNMs through the WKB method, and finally, we summarize our results and draw concluding remarks in Section \ref{concluding}. Throughout this work, we have used natural units $\hbar=c=G=1$.

\section{Self-Dual Black Hole}
\label{intr_sel_dual_bh}

In this section, we will briefly introduce the Self-Dual Black Hole solution that arises from a simplified model of LQG consisting of asymmetry reduced model corresponding to homogeneous spacetimes \cite{Modesto:2009ve}.

The structure of SDBH corresponds to a quantum version of the SchBH, and is described by the metric below:
\begin{equation} \label{lqgbh_metric}
 \begin{aligned}
  ds^2 &= - G(r)dt^2 + \frac{dr^2}{F(r)} + H(r)\left ( d\theta^2 + \text{sin}^2 \theta d\phi^2\right ) ,
 \end{aligned}
\end{equation}
where the functions $G(r)$, $F(r)$ and $H(r)$ are given by
\begin{equation} \label{funct_metric}
 \begin{aligned}
 G(r) = \frac{(r-r_{+})(r-r_{-})(r+r_{*})^2}{r^{4}+a_{0}^{2}}, \ \ F(r) = {\frac{(r-r_{+})(r-r_{-})r^{4}}{(r+r_{*})^{2}(r^{4}+a_{0}^{2})}}, \ \ H(r) = \left(r^{2} + \frac{a_{0}^{2}}{r^{2}} \right).
 \end{aligned}
\end{equation}
In the functions \eqref{funct_metric}, we have the  presence of an external horizon localized in $r_{+}=2m$, an intermediate in $r_{*}=\sqrt{r_{+}r_{-}}$ and a Cauchy horizon localized at $r_{-}=2mP^2$, where the polymeric function $P$, is given by
\begin{equation} \label{polym_funct}
 P = \frac{\sqrt{1+\epsilon^{2}} - 1}{\sqrt{1+\epsilon^{2}} +1}.
\end{equation}
In the Eq. \eqref{polym_funct} we have the parameter $\epsilon=\gamma \delta_b$, where $\gamma$ is the Barbero-Immirzi parameter, and $\delta_b$ is polymeric parameter used for the quantization in LQG. Also, in the functions of the Eq. \eqref{funct_metric} appear the parameter $a_0$ defined by
\begin{equation}
a_{0} = \frac{A_{\text{min}}}{8\pi},
\end{equation}
where  $A_{\text{min}}$ is minimal area in context of LQG.

It is important to notice, that the Eq. \eqref{lqgbh_metric} is written in terms of the SDBH mass, where $m$ is associated with the ADM mass as follows
\begin{equation}
M = m(1 + \mathcal{P})^{2}.
\end{equation}
Furthermore, it is important to see that the function $H(r)$ defined in the Eq. \eqref{funct_metric} is equal to $r^2$ only in asymptotic limit. In this way, a new  physical radial coordinate that measures the circumference distance is given by
\begin{equation} \label{rad_coord}
 R = \sqrt{r^2 + \frac{a_0^2}{r^2}}.
\end{equation}

From Eq. \eqref{rad_coord}, we can see an important characteristic of the internal structure of the SDBH. When $r$ decreases from infinity to zero, the $R$ coordinate decreases from infinity to $R=\sqrt{2a_0}$ in $r=\sqrt{a_0}$, and then increases again to infinity. Considering the Eq. \eqref{rad_coord} in the external event horizon, i.e., in $r=r_{+}$, we get
\begin{equation}
 R_{+} = \sqrt{H(r_{+})} = \sqrt{(2m)^{2} + \Big(\frac{a_{0}}{2m}\Big)^{2}} . \label{r-horizon}
\end{equation}

\begin{figure}[h!]
\centering
\includegraphics[scale=0.7]{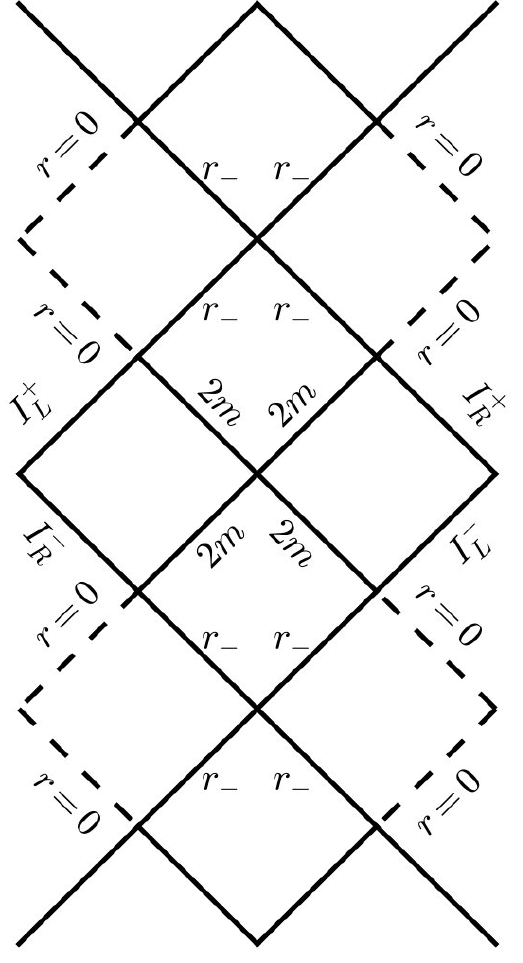}
\caption{Carter-Penrose diagram to the SDBH metric. The diagram has two asymptotic regions, being one at infinity and the other near the origin, where none observer can reach considering a finite time.}
\label{diag_bngql}
\end{figure}

A very interesting characteristic of this scenery is the self-duality of metric in Eq. \eqref{lqgbh_metric}. The self-duality means that, if we introduce new coordinates, $\tilde{r}=a_{0}/r$ and $ \tilde{t}=tr_{*}^{2}/a_{0}$, the form of metric is preserved. The dual radial coordinate is given by $\tilde{r}=\sqrt{a_0}$ and corresponds to a minimal element of surface. Furthermore, Eq. \eqref{rad_coord} can be written in the form $R=\sqrt{r^2+\tilde{r}^2}$  that clearly shows an asymptotically flat space, that is, a Schwarzschild region in the place of singularity to the limit when $r$ tends to zero. This region corresponds to a wormhole with the size of the order of the Planck length. The Carter-Penrose diagram for the SDBH is shown in Fig. \ref{diag_bngql}.

We can interpret the SDBH solution of the Eq. \eqref{lqgbh_metric} as a solution of the Einstein equation \cite{Modesto:2009ve}, given by
\begin{equation} \label{einst_eq}
 G_{\mu \nu} = R_{\mu \nu} - \frac{1}{2}g_{\mu \nu} R = 8 \pi T_{\mu \nu}^{\text{eff}},
\end{equation}
where $T_{\mu \nu}^{\text{eff}}$ is associated with an effective matter fluid that simulates the LQG corrections, given by
\begin{equation} \label{energy_tens}
 T_{\mu \nu}^{\text{eff}} = \text{diag}\left(-\rho,p_r,p_{\theta},p_{\theta}\right),
\end{equation}
where the components of the effective stress-energy tensor are: $\rho=-G_{tt}/8\pi$, $p_r=G_{rr}/8\pi$ and $p_{\theta}=G_{\theta \theta}/8\pi$. So, in the limit where $P$ and $a_0$ tends to zero, the SDBH metric of Eq. \eqref{lqgbh_metric} give us the Schwarzschild solution $\left(g_{\mu \nu}^{\text{Sch}}\right)$ that satisfies $G_{\mu \nu}\left(g_{\mu \nu}^{\text{Sch}}\right)\equiv 0$.


\section{Polar gravitational perturbations and effective potential}
\label{effect_potent}

In this section, we will use the method used by Zerilli for finding a master equation considering the polar gravitational perturbation \cite{Zerilli:1970se, Zerilli:1971wd}. So, if small perturbations are introduced, the resulting spacetime metric can be written in the form:
\begin{equation} \label{pert_metric_g}
 \tilde{g}_{\mu \nu} = g_{\mu \nu} + h_{\mu \nu} \ ,
\end{equation}
where  $g_{\mu \nu}$ is the background metric \eqref{lqgbh_metric} and $h_{\mu \nu}$ are the spacetime perturbations, where $\left |h_{\mu \nu}\right | \ll \left |g_{\mu \nu}\right |$.

Considering the spherical symmetry of a black hole metric, the perturbation of the metric $h_{\mu \nu}$ can be composed by a sum of two parts, where the first is characterized by axial perturbations $h_{\mu \nu}^{\text{axial}}$ (which has been analyzed in \cite{Cruz:2015bcj} by considering the SDBH). The second part is the polar perturbation $h_{\mu \nu}^{\text{polar}}$, where we can separate in two independent terms, one depending on time and radial coordinates and, the other depending on angular coordinates through spherical harmonics.

Considering this context, the polar gravitational perturbations in the Regge-Wheeler gauge \cite{Zerilli:1970se} is given by
\begin{equation} 
 \label{polar_pert_p}
  \begin{aligned}
  h^{\text{pol}}_{\mu \nu} =    \begin{bmatrix}
  G(r) H_0(r) & H_1(r) & 0 & 0 \\
  H_1(r) & F(r)^{-1 }H_2(r) & 0 & 0 \\
  0 & 0 & H(r) K(r) & 0 \\
  0 & 0 & 0 & H(r) \text{sin}^2\theta K(r)
\end{bmatrix} e^{-i\omega t} P_{l}(\cos \theta) \ ,
 \end{aligned}
\end{equation}
where $H_0(r)$, $H_1(r)$, $H_2(r)$ and $K(r)$ are unknown functions for the polar perturbations. Thus, substituting the Eqs. \eqref{pert_metric_g} and \eqref{polar_pert_p} in  Eq. \eqref{einst_eq} we get a set of equations, given by
\begin{equation}
 \label{dk_eq2}
 \begin{aligned}
  \frac{d K(r)}{d r} &= -\frac{1}{2}\Bigg{[}\frac{H^{\prime}(r)}{H(r)}-\frac{G^{\prime}(r)}{G(r)}\Bigg{]}K(r) + \frac{1}{2}\frac{H^{\prime}(r)}{H(r)}Q(r) + \frac{i F(r)}{2 \omega}\Bigg{[}\frac{l(l+1)}{F(r)H(r)} -\frac{1}{2}\frac{G^{\prime}(r)}{G(r)} \\ & \ \ \ \times \frac{F^{\prime}(r)}{F(r)} - \frac{G^{\prime \prime}(r)}{G(r)} + \frac{1}{2}\frac{G^{\prime}(r)^2}{G(r)^2} - \frac{G^{\prime}(r)}{G(r)}\frac{H^{\prime}(r)}{H(r)}\Bigg{]}H_1(r) \ ,
 \end{aligned}
\end{equation}

\begin{equation}
 \begin{aligned}
  \frac{d Q(r)}{d r} = \frac{d K(r)}{d r} - \frac{i \omega}{G(r)}H_1(r) - \frac{G^{\prime}(r)}{G(r)}Q(r) \ , \ \ \ \ \ \ \ \ \ \ \ \ \ \ \ \ \ \ \ \ \ \ \ \ \ \ \ \ \ \ \ \ \ \ \ \ \ \ \ \ \ \ \ \ \ \ \ \ \
 \end{aligned}
\end{equation}

\begin{equation} \label{dh1_eq2}
 \begin{aligned}
  \frac{d H_1(r)}{d r} = -\frac{1}{2}\frac{F^{\prime}(r)}{F(r)}H_1(r) - \frac{1}{2}\frac{G^{\prime}(r)}{G(r)}H_1(r) - \frac{i \omega}{F(r)}Q(r) - \frac{i \omega}{F(r)}K(r) \ , \ \ \ \ \ \ \ \ \ \ \ \ \ \ \ \ \
 \end{aligned}
\end{equation}
where we have taken $H_0(r)=H_2(r)\equiv Q(r)$ and the prime means the derivative with respect to the radial coordinate. In addition to the equations above, the functions $K(r)$, $Q(r)$ and $H_1(r)$ also should satisfy the following algebraic identity:
\begin{equation}
 \label{recor_eq}
 \Lambda_0 Q(r) + \Lambda_1 K(r) + \Lambda_2 H_1(r) = 0 \ ,
\end{equation}
where
\begin{equation}
 \begin{aligned}
  \Lambda_0 &= -\frac{1}{2}F(r)H^{\prime \prime}(r) - \frac{1}{4}F^{\prime}(r)H^{\prime}(r) + \frac{l(l+1)}{2} - \frac{1}{4}\frac{G^{\prime}(r)F^{\prime}(r)H(r)}{G(r)} - \frac{1}{2}\frac{G^{\prime \prime}(r)}{G(r)}F(r)H(r) \\ & \ \ \ + \frac{1}{4}\frac{G^{\prime}(r)^2}{G(r)^2}F(r)H(r) \ ,
 \end{aligned}
\end{equation}

\begin{equation}
 \begin{aligned}
  \Lambda_1 &= \omega^2 \frac{H(r)}{G(r)} + \frac{1}{2}F(r)H^{\prime \prime}(r) - \frac{1}{2}l(l+1) + \frac{1}{4}F^{\prime}(r)H^{\prime}(r) + \frac{1}{4}\frac{G^{\prime}(r)^2F(r)H(r)}{G(r)^2} \ , \ \ \ \ \ \ \ \ 
  \end{aligned}
\end{equation}
and
\begin{equation}
 \begin{aligned}
  \Lambda_2 &= -\frac{1}{2}i\omega \frac{F(r)H^{\prime}(r)}{G(r)} - \frac{1}{2}\frac{iF(r)^2H(r)G^{\prime \prime \prime}(r)}{\omega G(r)} - \frac{3}{8}\frac{iF(r)^2H(r)G^{\prime}(r)^3}{\omega G(r)^3} \\ & \ \ \ + \frac{1}{2}\frac{iF(r)^2G^{\prime}(r)^2H^{\prime}(r)}{\omega G(r)^2} - \frac{1}{2}\frac{iF(r)^2G^{\prime}(r)H^{\prime \prime}(r)}{\omega G(r)} + \frac{3}{4}\frac{iF(r)^2H(r)G^{\prime}(r)G^{\prime \prime}(r)}{\omega G^2(r)} \\ & \ \ \ -\frac{iF(r)^2H^{\prime}(r)G^{\prime \prime}(r)}{\omega G(r)} - \frac{3}{4}\frac{iF(r)H(r)F^{\prime}(r)G^{\prime \prime}(r)}{\omega G(r)} - \frac{1}{4}\frac{iF(r)H(r)G^{\prime}(r)F^{\prime \prime}(r)}{\omega G(r)} \\ & \ \ \ +\frac{3}{8}\frac{iF(r)H(r)G^{\prime}(r)^2F^{\prime}(r)}{\omega G(r)^2} -\frac{3}{4}\frac{iF(r)H^{\prime}(r)G^{\prime}(r)F^{\prime}(r)}{\omega G(r)} + \frac{1}{4}\frac{iF(r)G^{\prime}(r)}{\omega G(r)}l(l+1) \ . \ \ \ \ 
 \end{aligned}
\end{equation}

Now, we will define a new function, $R(r) \equiv H_1(r)/\omega$, and so the Eq. \eqref{recor_eq} can be rewritten as
\begin{equation}
 Q(r) = \Big[\alpha(r)+\beta(r)\omega^2\Big]K(r) + \Big[\gamma(r)+\delta(r)\omega^2\Big]R(r) \ ,
\end{equation}
where
\begin{equation}
 \begin{aligned}
  \alpha(r) &= \Bigg{[}-F(r)H^{\prime \prime}(r)-\frac{1}{2}F^{\prime}(r)H^{\prime}(r)+l(l+1)-\frac{1}{2}\frac{G^{\prime}(r)F^{\prime}(r)H(r)}{G(r)}  -\frac{G^{\prime \prime}(r)}{G(r)}F(r) \\ & \ \ \ \ \times H(r)+\frac{1}{2}\frac{G^{\prime}(r)^2}{G(r)^2}F(r)H(r)\Bigg{]}^{-1} \Bigg{[}-F(r)H^{\prime \prime}(r)  +l(l+1)-\frac{1}{2}F^{\prime}(r)H^{\prime}(r) \\ & \ \ \ \ -\frac{1}{2}\frac{G^{\prime}(r)^2F(r)H(r)}{G(r)^2}\Bigg{]} \ ,
 \end{aligned}
\end{equation}

\begin{equation}
 \begin{aligned}
  \beta(r) &= \Bigg{[}-F(r)H^{\prime \prime}(r)-\frac{1}{2}F^{\prime}(r)H^{\prime}(r)+l(l+1)-\frac{1}{2}\frac{G^{\prime}(r)F^{\prime}(r)H(r)}{G(r)}  -\frac{G^{\prime \prime}(r)}{G(r)}F(r) \\ & \ \ \ \ \times H(r) +\frac{1}{2}\frac{G^{\prime}(r)^2}{G(r)^2}F(r)H(r)\Bigg{]}^{-1} \Bigg{[}-2\frac{H(r)}{G(r)}\Bigg{]} \ ,
 \end{aligned}
\end{equation}

\begin{equation}
 \begin{aligned}
  \gamma(r) &= \Bigg{[}-F(r)H^{\prime \prime}(r)-\frac{1}{2}F^{\prime}(r)H^{\prime}(r)+l(l+1)-\frac{1}{2}\frac{G^{\prime}(r)F^{\prime}(r)H(r)}{G(r)}  -\frac{G^{\prime \prime}(r)}{G(r)}F(r) \\ & \ \ \ \ \times H(r)+\frac{1}{2}\frac{G^{\prime}(r)^2}{G(r)^2}F(r)H(r)\Bigg{]}^{-1} \Bigg{[}\frac{iF(r)^2H(r)G^{\prime \prime \prime}(r)}{G(r)} +\frac{3}{4}\frac{iF(r)^2H(r)G^{\prime}(r)^3}{G(r)^3} \\ & \ \ \ \ -\frac{iF(r)^2G^{\prime}(r)^2H^{\prime}(r)}{G(r)^2}+\frac{iF(r)^2G^{\prime}(r)H^{\prime \prime}(r)}{G(r)} -\frac{3}{2}\frac{iF(r)^2H(r)G^{\prime}(r)G^{\prime \prime}(r)}{G(r)^2}+2 \\ & \ \ \ \ \times \frac{iF(r)^2H^{\prime}(r)G^{\prime \prime}(r)}{G(r)}+\frac{3}{2}\frac{iF(r)H(r)F^{\prime}(r)G^{\prime \prime}(r)}{G(r)} +\frac{1}{2}\frac{iF(r)H(r)G^{\prime}(r)F^{\prime \prime}(r)}{G(r)} \\ & \ \ \ \ -\frac{3}{4}\frac{iF(r)H(r)G^{\prime}(r)^2F^{\prime}(r)}{G(r)^2}+\frac{3}{2}\frac{iF(r)H^{\prime}(r)G^{\prime}(r)F^{\prime}(r)}{G(r)} -\frac{1}{2}\frac{iF(r)G^{\prime}(r)}{G(r)}l(l+1)\Bigg{]}
 \end{aligned}
\end{equation}
and
\begin{equation}
 \begin{aligned}
  \delta(r) &= \Bigg{[}-F(r)H^{\prime \prime}(r)-\frac{1}{2}F^{\prime}(r)H^{\prime}(r)+l(l+1)-\frac{1}{2}\frac{G^{\prime}(r)F^{\prime}(r)H(r)}{G(r)}  -\frac{G^{\prime \prime}(r)}{G(r)}F(r) \\ & \ \ \ \ \times H(r)+\frac{1}{2}\frac{G^{\prime}(r)^2}{G(r)^2}F(r)H(r)\Bigg{]}^{-1} \Bigg[i\frac{F(r)H^{\prime}(r)}{G(r)}\Bigg] \ .
 \end{aligned}
\end{equation}
Also, by substituting in the Eq. \eqref{dk_eq2}, we get
\begin{equation}
 \begin{aligned}
  \frac{dK(r)}{dr} = - \big[\alpha_0(r)+\alpha_2(r)\omega^2\Big]K(r) - \Big[\beta_0(r)+\beta_2(r)\omega^2\Big]R(r) ,
 \end{aligned}
\end{equation}
where
\begin{equation}
 \begin{aligned}
  \alpha_0(r) = \frac{1}{2}\Bigg[\frac{H^{\prime}(r)}{H(r)}-\frac{G^{\prime}(r)}{G(r)}\Bigg]-\frac{1}{2}\frac{H^{\prime}(r)}{H(r)}\alpha(r) \ , \ \ \ \alpha_2(r) = -\frac{1}{2}\frac{H^{\prime}(r)}{H(r)}\beta(r) \ ,
 \end{aligned}
\end{equation}

\begin{equation}
 \begin{aligned}
  \beta_0(r) &= -\frac{1}{2}\frac{H^{\prime}(r)}{H(r)}\gamma(r)-\frac{iF(r)}{2}\Bigg[\frac{l(l+1)}{F(r)H(r)}-\frac{1}{2}\frac{G^{\prime}(r)F^{\prime}(r)}{G(r)F(r)}-\frac{G^{\prime \prime}(r)}{G(r)} \\ & +\frac{1}{2}\frac{G^{\prime}(r)^2}{G(r)^2}-\frac{G^{\prime}(r)F^{\prime}(r)}{G(r)H(r)}\Bigg] \ , \ \ \ \beta_2(r) = -\frac{1}{2}\frac{H^{\prime}(r)}{H(r)}\delta(r) \ .
 \end{aligned}
\end{equation}
Now for the Eq. \eqref{dh1_eq2}, we get
\begin{equation}
 \frac{dR(r)}{dr} = -\Big[\gamma_0(r)+\gamma_2(r)\omega^2\Big]K(r)-\Big[\delta_0(r)+\delta_2(r)\omega^2\Big]R(r) \ ,
\end{equation}
where
\begin{equation}
 \begin{aligned}
  \gamma_0(r) = \frac{i}{F(r)}\Big[1+\alpha(r)\Big] \ , \ \ \ \gamma_2(r) = \frac{i}{F(r)}\beta(r) \ ,
 \end{aligned}
\end{equation}
\begin{equation}
 \begin{aligned}
  \delta_0(r) = \frac{i}{F(r)}\gamma(r)+\frac{1}{2}\Bigg[\frac{F^{\prime}(r)}{F(r)}+\frac{G^{\prime}(r)}{G(r)}\Bigg] \ , \ \ \ \delta_2(r) = \frac{i}{F(r)}\delta(r) \ .
 \end{aligned}
\end{equation}

Finally, to obtain a Schrodinger-type equation, we will perform a change of variables, given by
\begin{equation}
 \label{chang_vari}
 \begin{aligned}
  K(r) = f(r)\hat{K}(x) + g(r)\hat{R}(x) \ , \ \ \ R(r) = h(r)\hat{K}(x) + t(r)\hat{R}(x) \ ,
 \end{aligned}
\end{equation}
where
\begin{equation}
 \begin{aligned}
  g(r) = 1 \ \ , \ \ t(r) = -\frac{\alpha_2(r)}{\beta_2(r)} \ \ , \ \ f(r) = -\sqrt{G(r)F(r)}\Bigg[\alpha_0(r)-\beta_0(r)\frac{\alpha_2(r)}{\beta_2(r)}\Bigg] \
 \end{aligned}
\end{equation}
and
\begin{equation}
 \begin{aligned}
  h(r) = \frac{1}{\beta_2(r)}\Bigg{\{}\frac{1}{\sqrt{G(r)F(r)}}+\sqrt{G(r)F(r)}\Bigg[\alpha_0(r)\alpha_2(r)-\beta_0(r)\frac{\alpha_2(r)^2}{\beta_2(r)}\Bigg]\Bigg{\}} \ .
 \end{aligned}
\end{equation}
Together with the Eqs. \eqref{chang_vari}, we still need to impose:
\begin{equation} \label{func_cond}
 \begin{aligned}
  \frac{d\hat{K}(x)}{dx} = \hat{R}(x) \ \ \ \text{and} \ \ \ \frac{d\hat{R}(x)}{dx} = \Big[V_{\text{eff}}(r) - \omega^2\Big]\hat{K}(x) \ ,
 \end{aligned}
\end{equation}
where the variable $x$ is called tortoise coordinate and is given by
\begin{equation}
 \begin{aligned}
  \frac{dr}{dx} = \sqrt{G(r)F(r)} \ .                                                                                                 
 \end{aligned}
\end{equation}
Consequently, from Eq. \eqref{func_cond} we get a Schr\"odinger-type equation given by
\begin{equation} \label{schr_typ_eq}
 \begin{aligned}
  \frac{d^2\hat{K}(x)}{dx^2} + \Big[\omega^2 - V_{\text{eff}}\big(r(x)\big)\Big]\hat{K}(x) = 0 \ ,
 \end{aligned}
\end{equation}
where the effective potential, $V_{\text{eff}}(r)$, is given by
\begin{equation} \label{effec_potent}
 \begin{aligned}
  V_{\text{eff}}(r) = -\sqrt{G(r)F(r)}\Bigg[\alpha_0(r)f(r)+\beta_0(r)h(r)+\frac{df(r)}{dr}\Bigg] \ .
 \end{aligned}
\end{equation}

\begin{figure}[h!]
    \centering
    \subfloat[]
    {{\includegraphics[width=7.5cm]{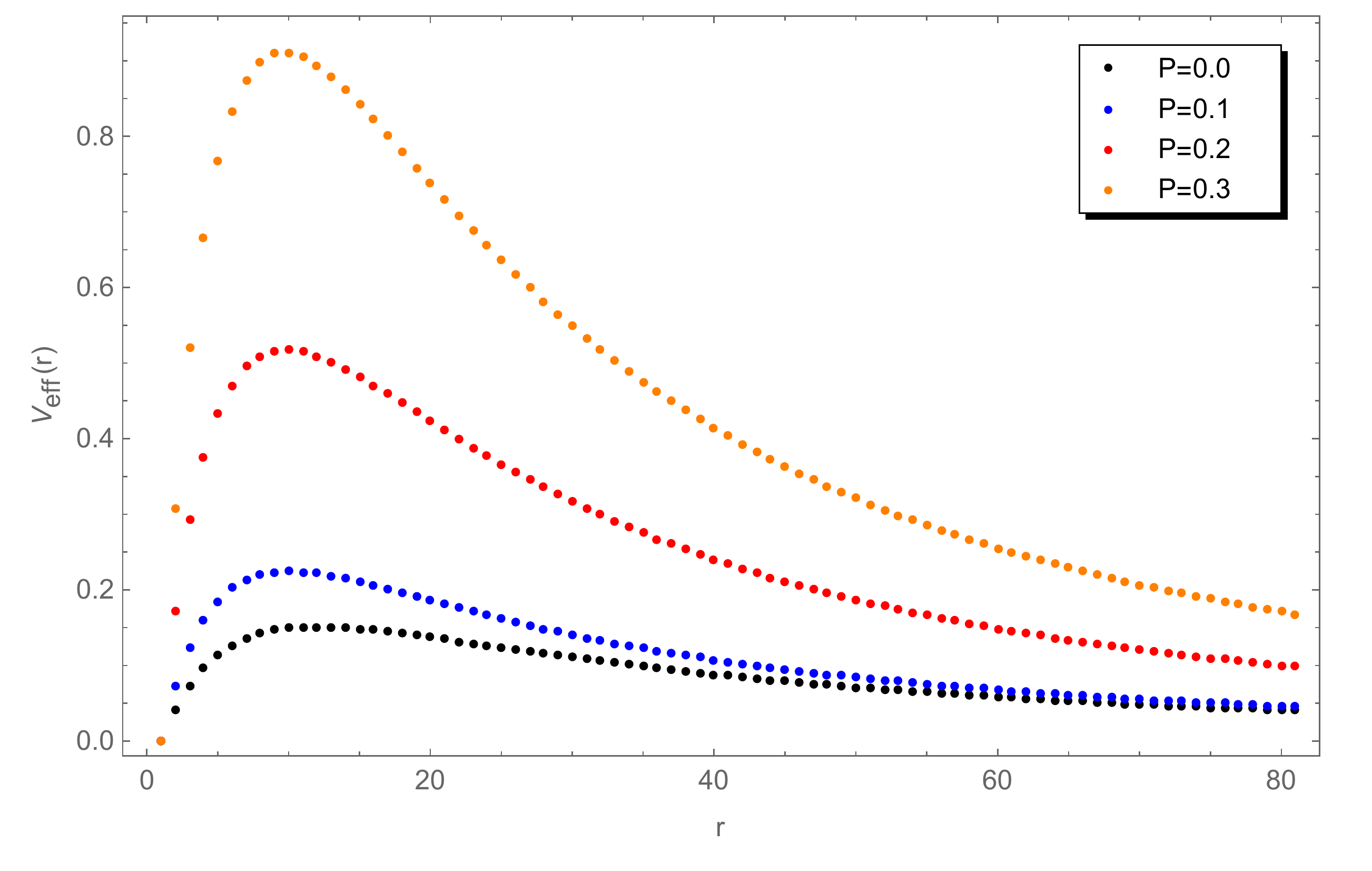} }}
    \qquad
    \subfloat[]
    {{\includegraphics[width=7.5cm]{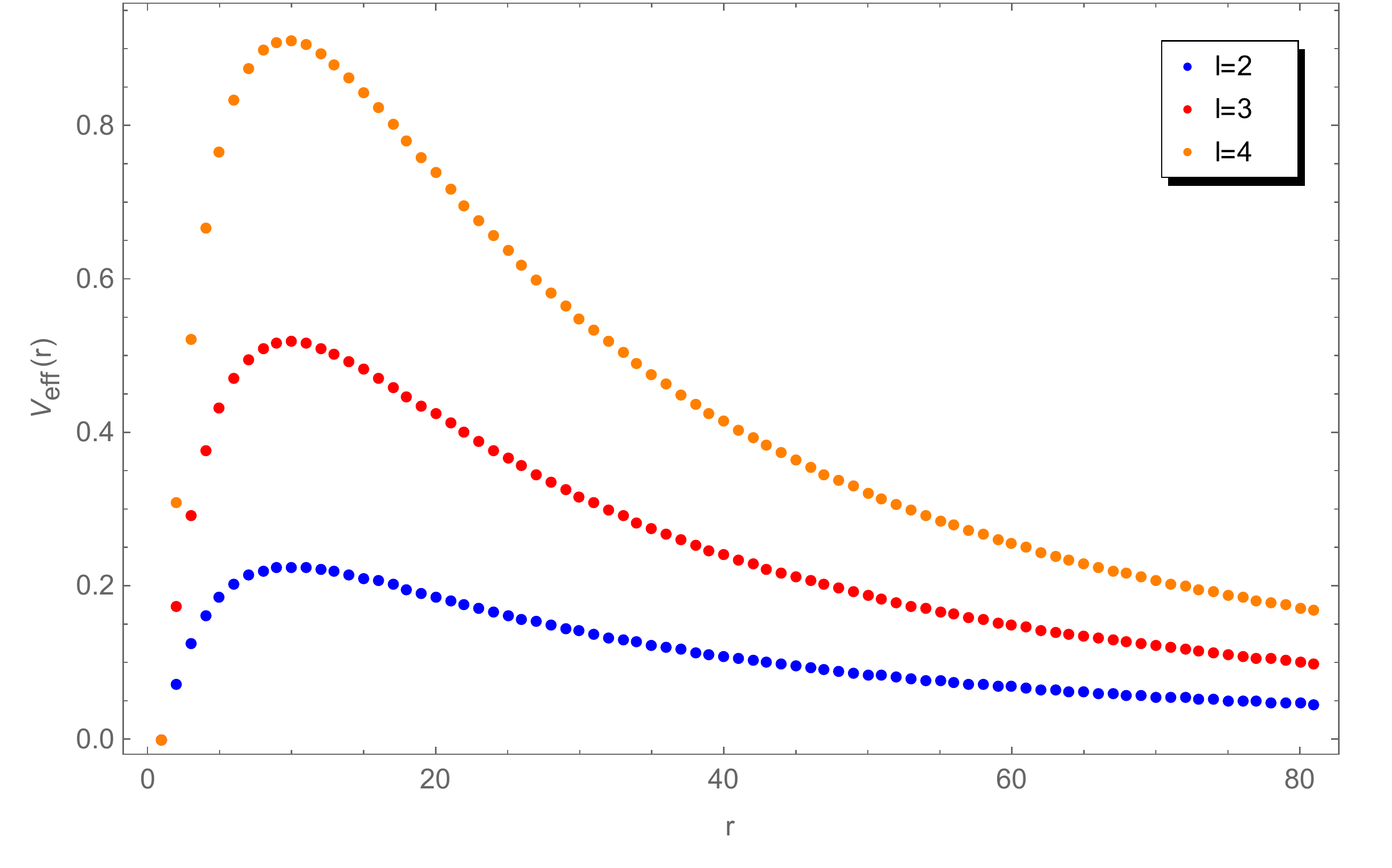}}}
    \caption{The behavior of the effective potential for polar gravitational perturbations. In both graphs, we have considered the BH mass $m=1$ and $a_0=\sqrt{3}/2$. In the plot (a) we consider the multipole value $l=2$ and vary the polymeric parameter as $P=0.0, 0.1, 0.2 \ \text{and} \ 0.3$. Here, the polymeric parameter value $P=0.0$ (and also $a_0=0$) are associated with the case of SchBH. In the plot (b) we consider the fixed value $P=0.1$ and vary the multipole values as $l=2, 3 \ \text{and} \ 4$.}
    \label{effec_potent_graph}
\end{figure}

The behavior of the effective potential of the Eq. \eqref{effec_potent} is shown in the Fig. \ref{effec_potent_graph} considering different values for the polymeric parameter $P$ and for the multipole number $l$. We can see that, when $P$ and $a_0$ tends to zero, the effective potential tends to the classical Zerilli potential for the SchBH \cite{Zerilli:1970se}.

\section{Quasinormal modes} 
\label{qnm_sdbh}

In this section, our focus will be the calculus of QNMs of SDBH described by the metric of the Eq. \eqref{lqgbh_metric}. As we saw in the last section, the polar perturbations to the SDBH metric and, consequently, the field equations can be reduced to a Schrodinger-type wave equation given by Eq. \eqref{schr_typ_eq}. The effective potential $V_{\text{eff}}(x)$, that appears in Eq. \eqref{effec_potent} is constant in the event horizon ($x=-\infty$) and at infinity ($x=\infty$) and has a maximum at some point intermediate ($x=x_0$).

Thus, we can make a direct analogy with the problem of scattering near the peak of the barrier potential in Quantum Mechanics (QM), where $\omega^2$ in Eq. \eqref{schr_typ_eq} plays the role of the energy. Several methods to find the QNMs have been developed, however, we chose an approximated method, which is the well-known WKB approach introduced by Schutz and Will \cite{Schutz:1985km}. This treatment was later improved to the 3rd order by Iyer and Will \cite{Iyer:1986np} and is the approximation that we shall consider. For a review of the available QNMs techniques, we suggest \cite{Berti:2009kk} and \cite{Dreyer:2002vy, Konoplya:2004ip, Cardoso:2003cj, Kokkotas:1999bd, Nollert:1999ji} for further readings.

Therefore, with the objective of following the WKB method, we have supposed that $\hat{K}(x(r))$ have a harmonic asymptotic behavior in $t$ coordinate, $\hat{K}(x) \sim e^{-i\omega(t\pm x)}$ and $V_{\text{eff}}(r(x)) \rightarrow 0$ when $x \rightarrow \pm \infty$ in Eq. \eqref{schr_typ_eq}. Thus, the QNMs $\omega_{n}$ that appear in Eq. \eqref{schr_typ_eq} are determined (up to 3rd order of the WKB approximation), by the following equation:
\begin{equation} 
 \label{qnm_eq}
 \omega_{n}=\sqrt{\left(V_{0} + \Delta \right) - i \left(n + \frac{1}{2}\right) \sqrt{-2 V^{\prime \prime}_{0}} \left(1 + \Omega \right)} \ ,
\end{equation}
where
\begin{equation}
 \Delta = \frac{1}{8}\left(\frac{V_{0}^{(4)}}{V^{\prime \prime}_{0}}\right)
 \left(\frac{1}{4}+\alpha^{2}\right)-\frac{1}{288}\left(\frac{V^{\prime \prime \prime}_{0}}{V^{\prime \prime}_{0}}\right)^{2}\left(7+60\alpha^{2}
 \right) \ ,
\end{equation}
\begin{equation}
 \begin{aligned}
 \Omega &= -\frac{1}{2V^{\prime \prime}_{0}} \left \{ \frac{5}{6912}\left(\frac{V^{\prime \prime \prime}_{0}}{V^{\prime \prime}_{0}}\right)^{4}\left(77+188\alpha^{2}
 \right) - \frac{1}{384}\left[\frac{\left(V^{\prime \prime \prime}_{0}\right)^{2}\left(V^{(4)}_{0}\right)}{\left(V^{\prime \prime}_{0}\right)^{3}}\right]\left(51+100\alpha^{2}
 \right) \right. \\ & \ \ \ \left. \right. + \frac{1}{2304}\left(\frac{V^{(4)}_{0}}{V^{\prime \prime}_{0}}\right)^{2}\left(65+68\alpha^{2}\right) + 
 \frac{1}{288}\left(\frac{V^{\prime \prime \prime}_{0}V^{(5)}_{0}}{\left(V^{\prime \prime}_{0}\right)^{2}}\right)\left(19+28\alpha^{2}\right) \\ & \ \ \ \left. - \frac{1}{288}\left(\frac{V^{(6)}_{0}}{V^{\prime \prime}_{0}}\right)\left(5+4\alpha^{2}\right) \right \} \ .
 \end{aligned}
\end{equation}
In the relations above, we have $\alpha=n+\frac{1}{2}$ and $V_0^{(n)}$ that denotes the $n$-order derivative of the effective potential on the point maximum $x_0$. Thus, using the Eq. \eqref{effec_potent} and Eq. \eqref{qnm_eq}, we can calculate the QNMs, $\omega_n$ for the SDBH, which are shown in the tables \ref{table1}, \ref{table2} and \ref{table3} for the different values of LQG parameter. For the case that the polymeric $P$ and minimal area $a_0$ parameters tends to zero, the results converge to SchBH \cite{Konoplya:2004ip}.

\begin{table}[!h]
\begin{center}
\scriptsize
\begin{tabular}{|c|c|c|c|c|}
\hline
{\bf $P$} & {\bf $\omega_{0}$} & {\bf $\omega_{1}$} & {\bf $\omega_{2}$} \\
\hline
{\bf $0.1$} & {\bf $0.4588080\, -0.0943413 i$} & {\bf $0.4350910\, -0.2893620 i$} & 
{\bf $0.3976210\, -0.4945600 i$} \\
\hline
{\bf $0.2$} & {\bf $0.5294850\, -0.0964412 i$} & {\bf $0.5129000\, -0.2947210 i$} & 
{\bf $0.4883820\, -0.5020030 i$}\\
\hline
{\bf $0.3$} & {\bf $0.5893080\, -0.0945579 i$} & {\bf $0.5760760\, -0.2862250 i$} &
{\bf $0.5534050\, -0.4828290 i$}\\
\hline
{\bf $0.4$} & {\bf $0.6373510\, -0.0912246 i$} & {\bf $0.6285820\, -0.2768290 i$} & 
{\bf $0.6158650\, -0.4686530 i$}\\
\hline
{\bf $0.5$} & {\bf $0.6724460\, -0.0848729 i$} & {\bf $0.6658430\, -0.2560750 i$} & 
{\bf $0.6546990\, -0.4304320 i$}\\
\hline
{\bf $0.6$} & {\bf $0.6928030\, -0.0769778 i$} & {\bf $0.6881010\, -0.2318850 i$} & 
{\bf $0.6799620\, -0.3889860 i$}\\
\hline
{\bf $0.7$} & {\bf $0.6962800\, -0.0677585 i$} & {\bf $0.6926040\, -0.2040160 i$} & 
{\bf $0.6861650\, -0.3420780 i$}\\
\hline
{\bf $0.8$} & {\bf $0.6804520\, -0.0584343 i$} & {\bf $0.6768360\, -0.1776110 i$} & 
{\bf $0.6723440\, -0.3025970 i$}\\
\hline
{\bf $0.9$} & {\bf $0.6441710\, -0.0499027 i$} & {\bf $0.6394600\, -0.1522590 i$} & 
{\bf $0.6328630\, -0.2612890 i$}\\
\hline
\end{tabular}
\caption{First QNMs of SDBH considering $a_0=\sqrt{3}/2$ and $l = 2$.}
\label{table1}
\end{center}
\end{table}

\begin{table}[!h]
\begin{center}
\scriptsize
\begin{tabular}{|c|c|c|c|c|}
\hline
{\bf $P$} & {\bf $\omega_{0}$} & {\bf $\omega_{1}$} & {\bf $\omega_{2}$} \\
\hline
{\bf $0.1$} & {\bf $0.7103050\, -0.0954021 i$} & {\bf $0.6957200\, -0.2872270 i$} & 
{\bf $0.6675730\, -0.4815990 i$} \\
\hline
{\bf $0.2$} & {\bf $0.8001510\, -0.0968958 i$} & {\bf $0.7892090\, -0.2921900 i$} & 
{\bf $0.7692650\, -0.4909660 i$}\\
\hline
{\bf $0.3$} & {\bf $0.8777030\, -0.0953165 i$} & {\bf $0.8693880\, -0.2869460 i$} &
{\bf $0.8539780\, -0.4809750 i$}\\
\hline
{\bf $0.4$} & {\bf $0.9407610\, -0.0915475 i$} & {\bf $0.9347890\, -0.2760760 i$} & 
{\bf $0.9246220\, -0.4641020 i$}\\
\hline
{\bf $0.5$} & {\bf $0.9870080\, -0.0852833 i$} & {\bf $0.9826610\, -0.2565200 i$} & 
{\bf $0.9747190\, -0.4294660 i$}\\
\hline
{\bf $0.6$} & {\bf $1.0134100\, -0.0773285 i$} & {\bf $1.0102700\, -0.2325430 i$} & 
{\bf $1.0045700\, -0.3892210 i$}\\
\hline
{\bf $0.7$} & {\bf $1.0165400\, -0.0679873 i$} & {\bf $1.0140600\, -0.2042550 i$} & 
{\bf $1.0093900\, -0.3413180 i$}\\
\hline
{\bf $0.8$} & {\bf $0.9926870\, -0.0581686 i$} & {\bf $0.9901220\, -0.1744740 i$} & 
{\bf $0.9849480\, -0.2906990 i$}\\
\hline
{\bf $0.9$} & {\bf $0.9395680\, -0.0497044 i$} & {\bf $0.9362930\, -0.1496610 i$} & 
{\bf $0.9301570\, -0.2511700 i$}\\
\hline
\end{tabular}
\caption{First QNMs of SDBH considering $a_0=\sqrt{3}/2$ and $l = 3$.}
\label{table2}
\end{center}
\end{table}

\begin{table}[!h]
\begin{center}
\scriptsize
\begin{tabular}{|c|c|c|c|c|}
\hline
{\bf $P$} & {\bf $\omega_{0}$} & {\bf $\omega_{1}$} & {\bf $\omega_{2}$} \\
\hline
{\bf $0.1$} & {\bf $0.9472730\, -0.0967026 i$} & {\bf $0.9372570\, -0.2922060 i$} & 
{\bf $0.9198500\, -0.4928240 i$} \\
\hline
{\bf $0.2$} & {\bf $1.0587100\, -0.0973907 i$} & {\bf $1.0507900\, -0.2929440 i$} & 
{\bf $1.0357700\, -0.4904810 i$}\\
\hline
{\bf $0.3$} & {\bf $1.1554600\, -0.0956521 i$} & {\bf $1.1493700\, -0.2872840 i$} &
{\bf $1.1375000\, -0.4797910 i$}\\
\hline
{\bf $0.4$} & {\bf $1.2344300\, -0.0918630 i$} & {\bf $1.2298300\, -0.2768920 i$} & 
{\bf $1.2219900\, -0.4653630 i$}\\
\hline
{\bf $0.5$} & {\bf $1.2925000\, -0.0855016 i$} & {\bf $1.2892200\, -0.2568970 i$} & 
{\bf $1.2830100\, -0.4293580 i$}\\
\hline
{\bf $0.6$} & {\bf $1.3254200\, -0.0774855 i$} & {\bf $1.3230500\, -0.2327260 i$} & 
{\bf $1.3185400\, -0.3887160 i$}\\
\hline
{\bf $0.7$} & {\bf $1.3285600\, -0.0681213 i$} & {\bf $1.3267200\, -0.2044660 i$} & 
{\bf $1.3231200\, -0.3410980 i$}\\
\hline
{\bf $0.8$} & {\bf $1.2969100\, -0.0583603 i$} & {\bf $1.2949200\, -0.1753430 i$} & 
{\bf $1.2911200\, -0.2930790 i$}\\
\hline
{\bf $0.9$} & {\bf $1.2275200\, -0.0496775 i$} & {\bf $1.2252700\, -0.1487480 i$} & 
{\bf $1.2205700\, -0.2469960 i$}\\
\hline
\end{tabular}
\caption{First QNMs of SDBH considering $a_0=\sqrt{3}/2$ and $l = 4$.}
\label{table3}
\end{center}
\end{table}

\begin{figure}[h!]
    \centering
    \subfloat[Real part]{{\includegraphics[width=7.0cm]{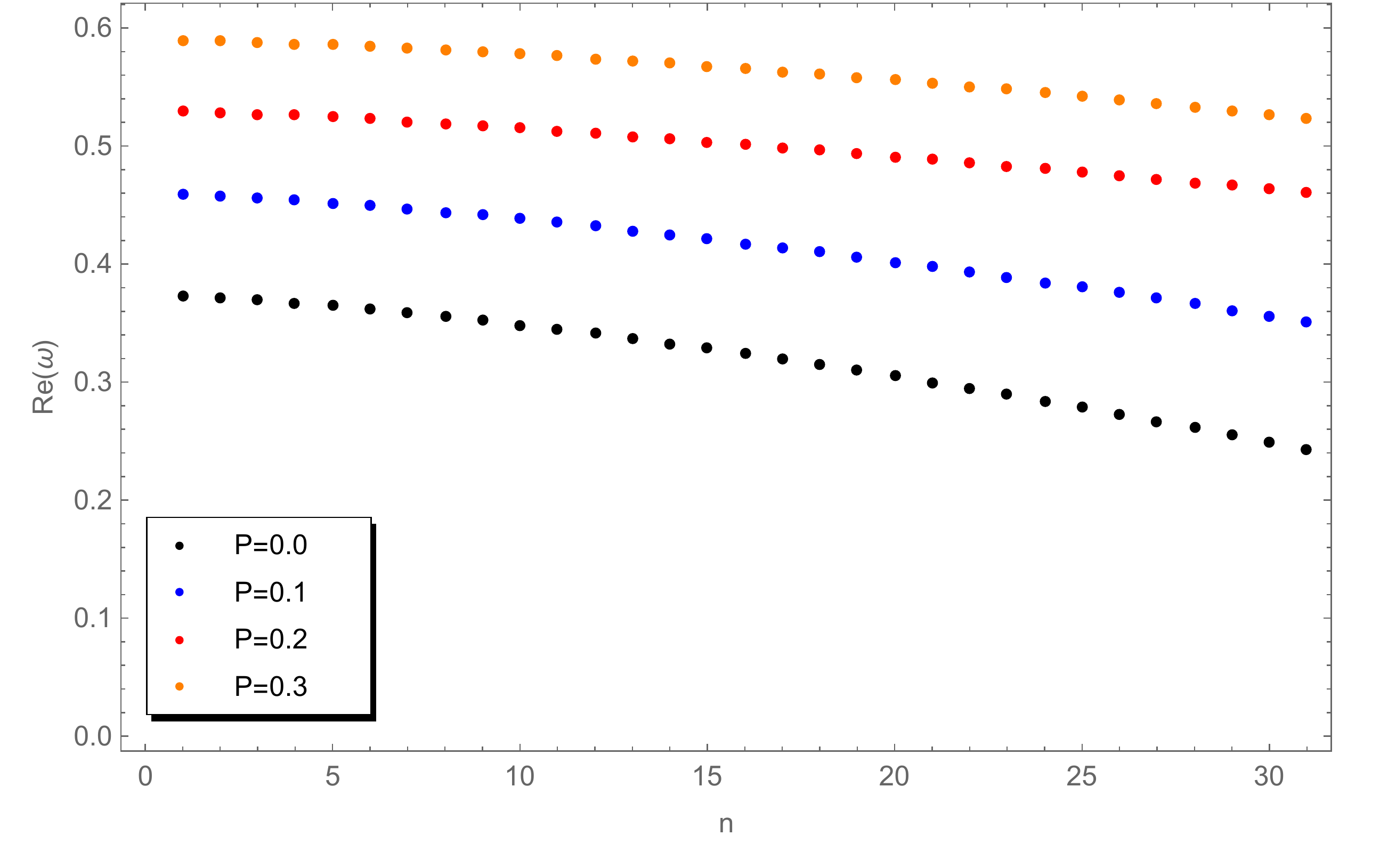} }}
    \qquad
    \subfloat[Imaginary part]{{\includegraphics[width=7.0cm]{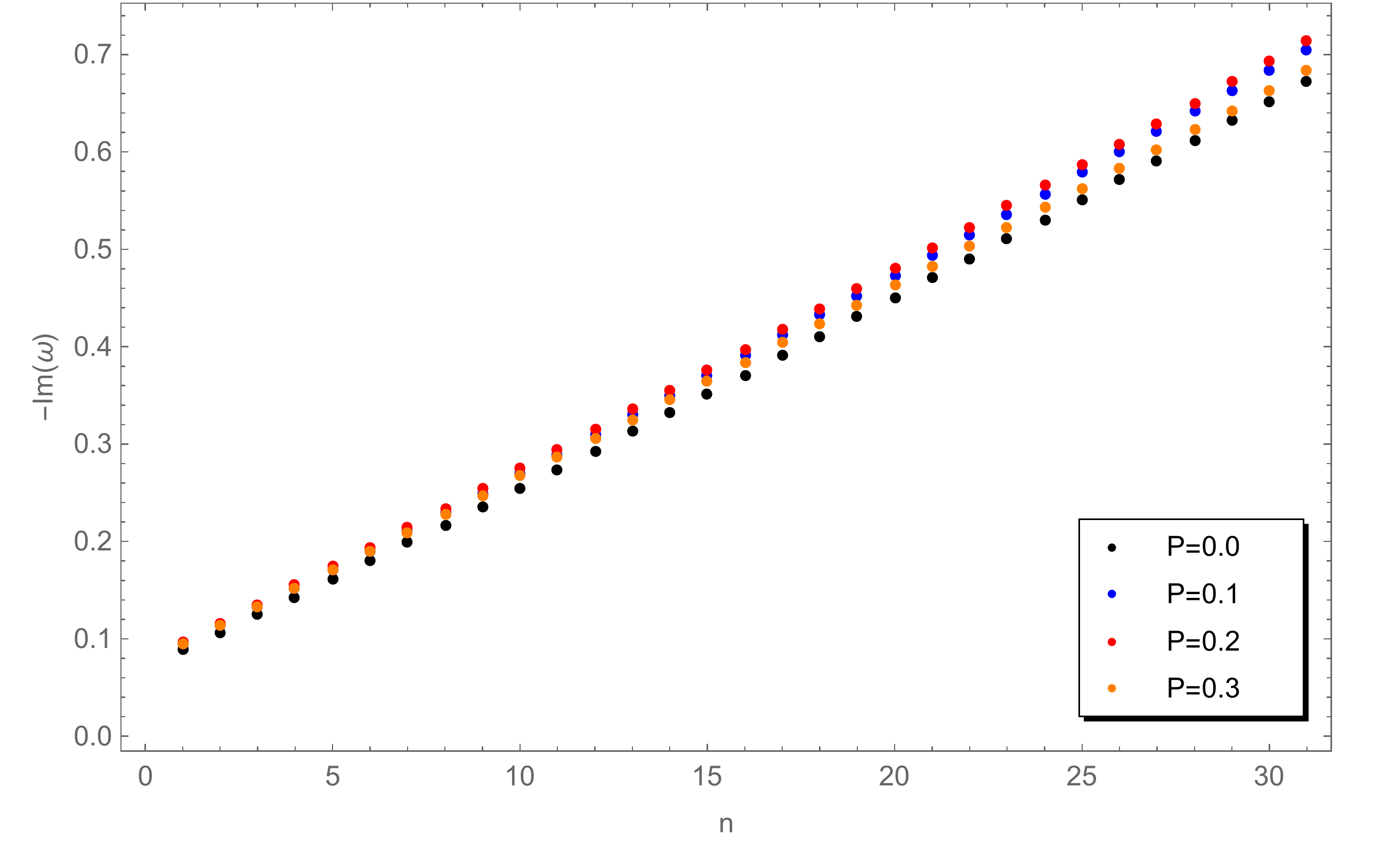} }}
    \caption{Graphs for the behavior of the QNMs considering $l=2$ and $P=0.0, 0.1, 0.2 \ \text{and} \ 0.3$. In plot (a) is shown the real part, while the imaginary part is shown in (b).}
    \label{graph_w_l2}
\end{figure}

\begin{figure}[h!]
    \centering
    \subfloat[Real part]{{\includegraphics[width=7.0cm]{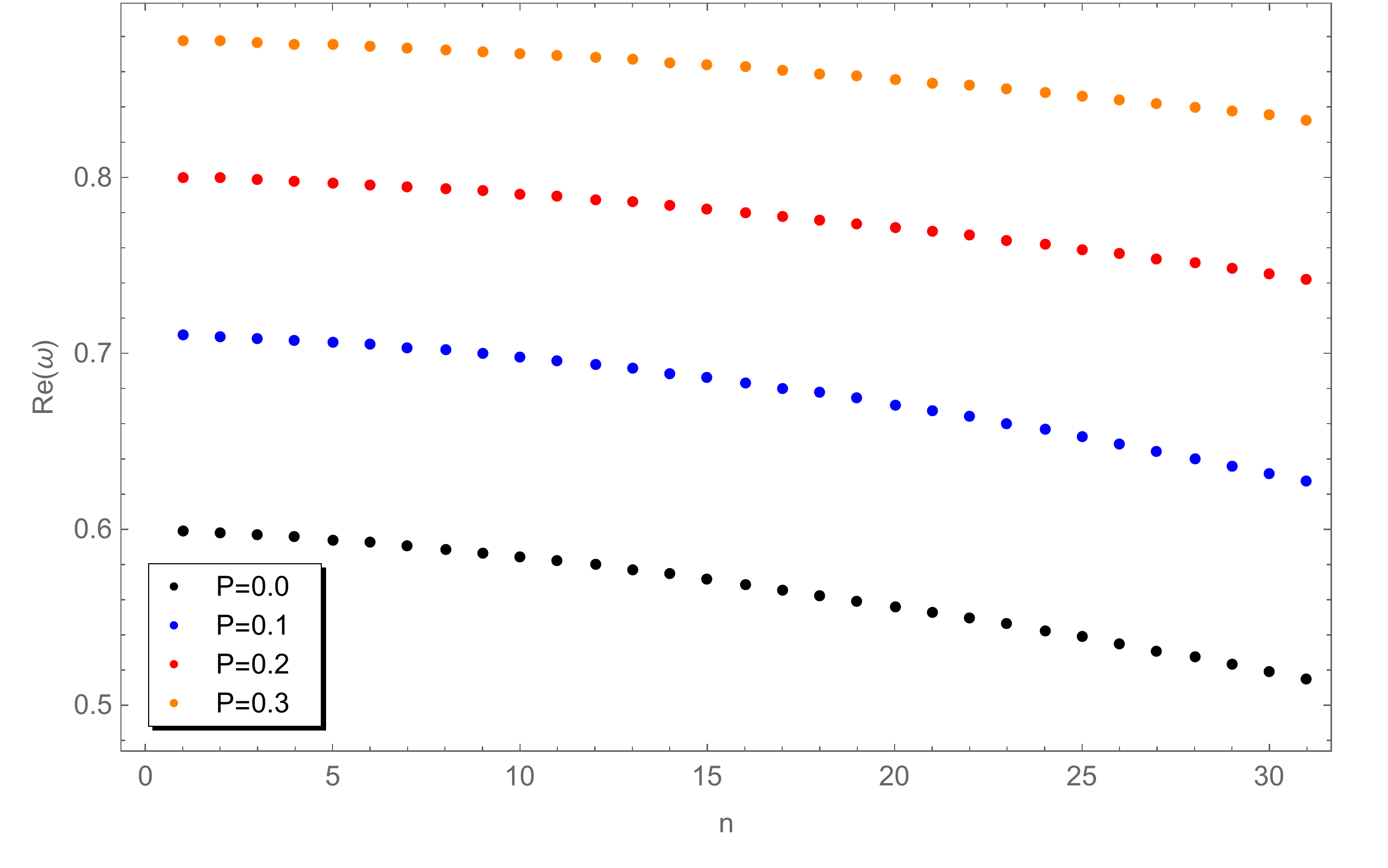} }}
    \qquad
    \subfloat[Imaginary part]{{\includegraphics[width=7.0cm]{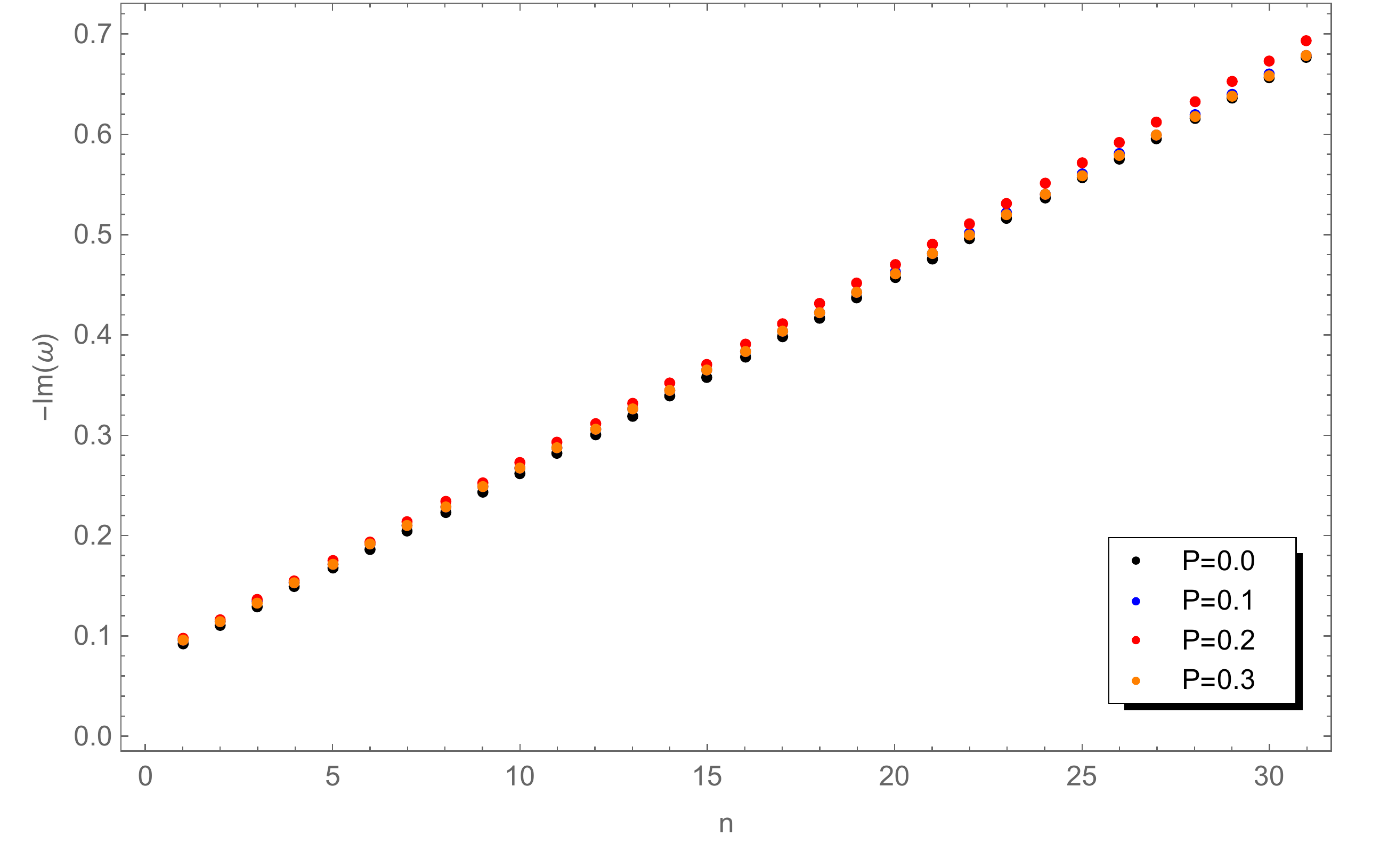} }}
    \caption{Graphs for the behavior of the QNMs considering $l=3$ and $P=0.0, 0.1, 0.2 \ \text{and} \ 0.3$. In plot (a) is shown the real part, while the imaginary part is shown in (b).}
    \label{graph_w_l3}
\end{figure}

\begin{figure}[h!]
    \centering
    \subfloat[Real part]{{\includegraphics[width=7.0cm]{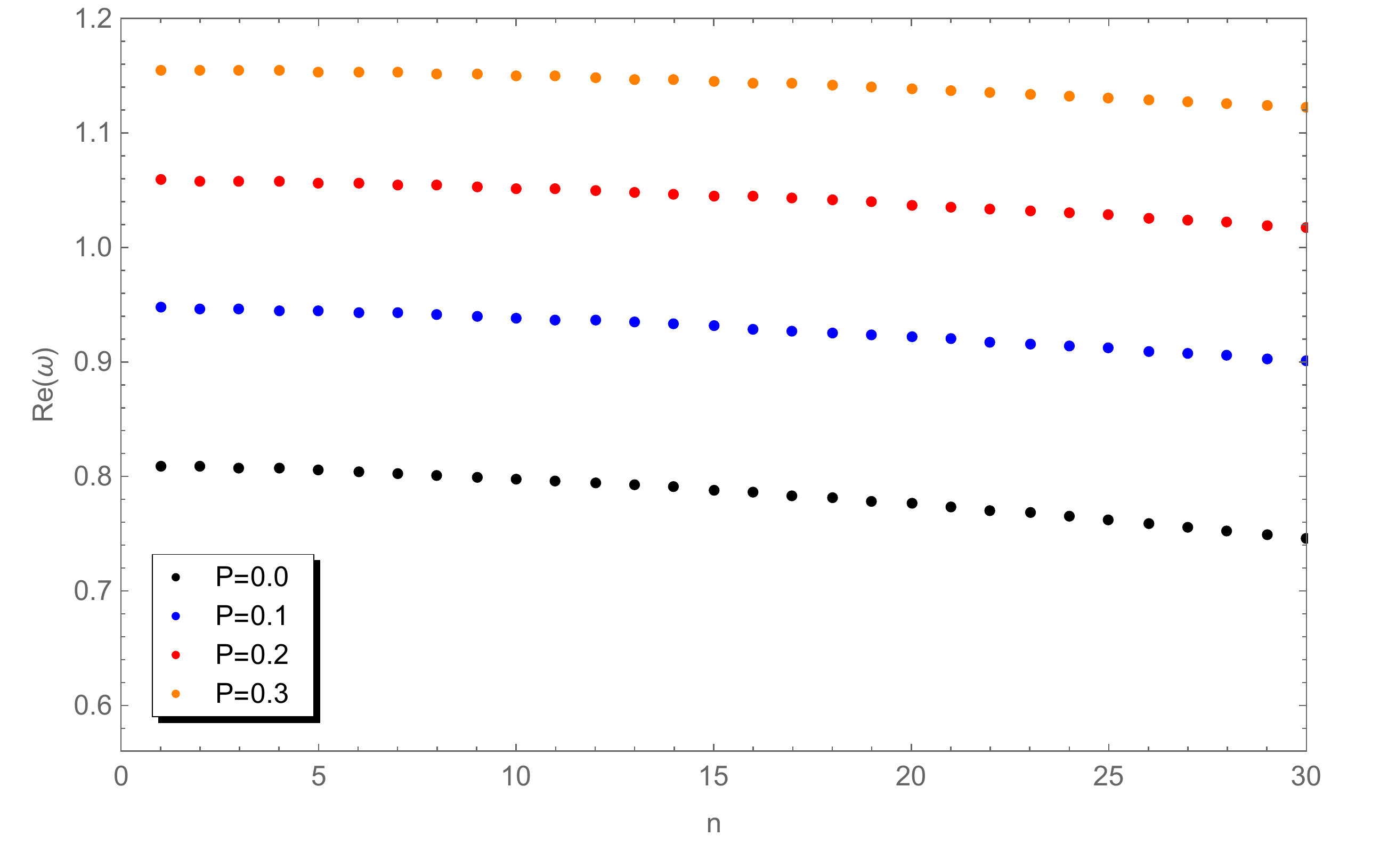} }}
    \qquad
    \subfloat[Imaginary part]{{\includegraphics[width=7.0cm]{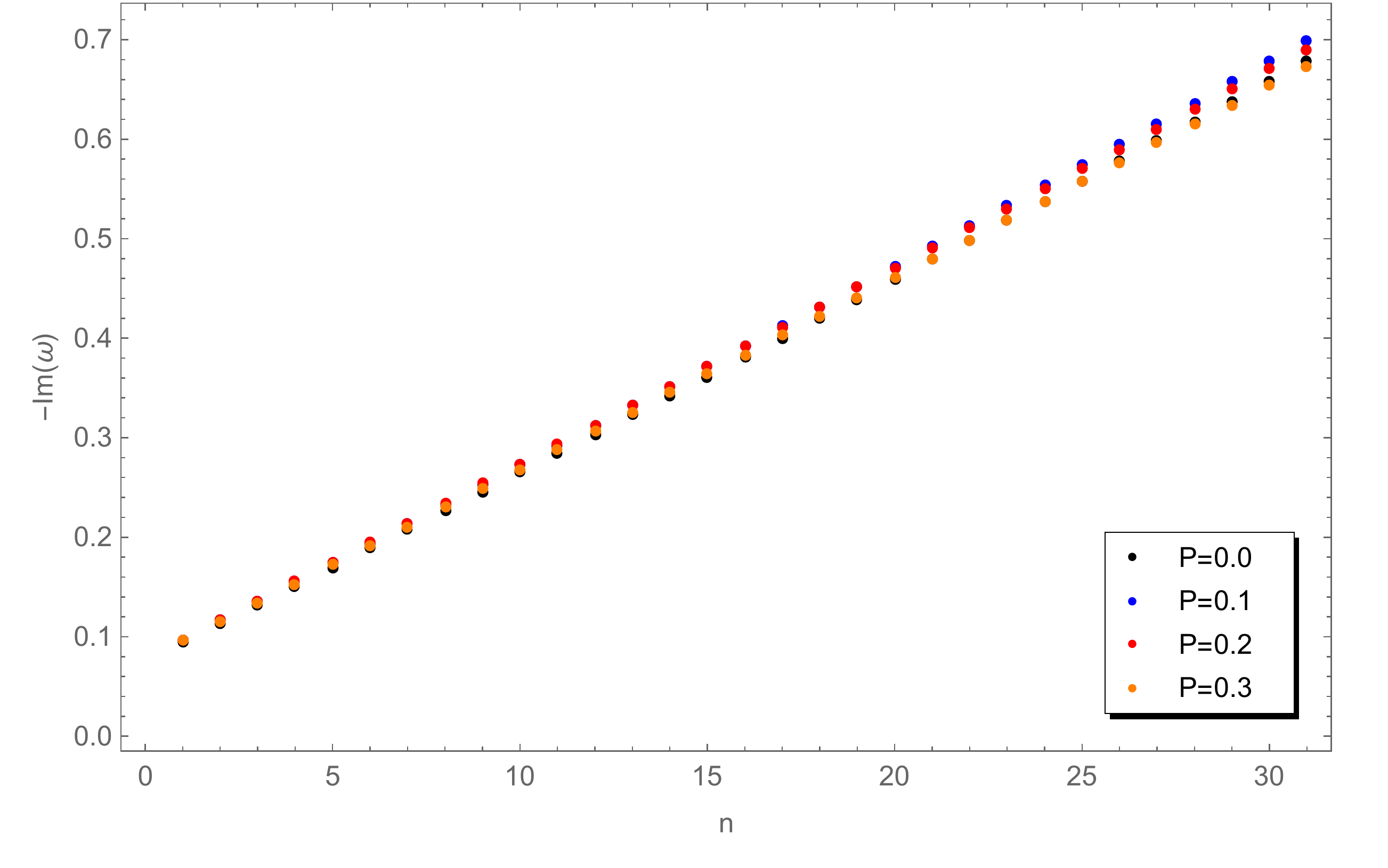} }}
    \caption{Graphs for the behavior of the QNMs considering $l=4$ and $P=0.0, 0.1, 0.2 \ \text{and} \ 0.3$. In plot (a) is shown the real part, while the imaginary part is shown in (b).}
    \label{graph_w_l4}
\end{figure}

For the best visualization of the effects of the LQG corrections in the QNMs spectrum, we have shown the behavior by graphs in Figs. \ref{graph_w_l2}, \ref{graph_w_l3}, and \ref{graph_w_l4}, as a function of $n$ and different values of the polymeric function $P$. It has been plotted the real and imaginary parts of $\omega$ for the $l=2, 3$ and $4$ and, considering the following values of the polymeric function: $P=0.1, 0.2 \ \text{and} \ 0.6$. In addition, for effect of comparison, it has been plotted together the behavior of the QNMs spectrum for the SchBH, i.e., $P=0$ and also $a_0=0$.

\section{Concluding remarks}
\label{concluding}

Gravitational-waves observations have opened a new window to gravitational physics research. In this framework, the black holes offer a great scenario to test the predictions of candidates to quantum gravity theories. With this in mind, we have studied the black hole perturbations and quasinormal modes spectrum to a quantized version of the Schwarzschild solution, which is known as a self-dual black hole \cite{Modesto:2009ve, Modesto:2008im}, and consists of a LQG solution.

Particularly, in the present work, we have considered small polar gravitational perturbations, Eq. \eqref{polar_pert_p}, which linearizes the Einstein equation, Eq. \eqref{einst_eq}, and provides a Schrodinger-type equation with an effective potential given by Eq. \eqref{effec_potent}. Next, we use the WKB approach to getting the quasinormal modes showed in the tables \ref{table1}, \ref{table2} and \ref{table3}. We consider different values to the polymeric parameter, going $0.1$ to $0.9$, and also we assumed the values $2$, $3$ and $4$, to angular number $l$. However, to a best visualization of polar gravitational QNMs behavior we ploted in Figs. \ref{graph_w_l2}, \ref{graph_w_l3} and \ref{graph_w_l4} the real and imaginary parts. For the graphs, we have chosen the values of $P=0.1, \ 0.2 \ \text{and} \ 0.3$ and $l=2, 3 \ \text{and} \ 4$.

So, analyzing our results, we can verify that the polar quasinormal modes strongly depend on the LQG parameters.
Our results show that as the parameter $P$ grows, the real part of the QNMs suffers an initial increase and then starts to decrease, while the magnitude of the imaginary part decreases, considering the fixed-parameter $a_0$. This behavior is verified also in \cite{Santos:2015gja, Cruz:2015bcj}. This characteristic reveals that the damping of polar perturbations in the self-dual black hole is slower and the oscillations are faster or slower depending on the value of $P$. Based on these results, we can conclude that the self-dual black hole has a stable behavior under polar gravitational perturbations.

On the other hand, by comparing our results with that obtained considering the axial gravitational perturbations \cite{Cruz:2015bcj}, for the same values of the LQG parameters, we have observed different values for the QNMs frequences. It points to the breaking of the isospectrality in the LQBH scenario. In this way, the present analysis based on LQG shows that isospectrality may not be held in the presence of quantum gravity corrections to the Schwarzschild metric.

So, as the present study together with \cite{Cruz:2015bcj} concludes the analysis on the linear stability of the self-dual black hole, it also opens the discussion about the interesting issue of isospectrality in the context of loop quantum black holes. Further analysis considering charged and rotating extensions of SDBH, as well as, other black hole solutions in LQG, must improve our understanding about this issue. Such studies can be relevant concerning gravitational wave astronomy, because the improvement of the sensitivities of the detectors, it could be possible for the LIGO-VIRGO collaboration to measure the QNMs.

However, it is in order to mention that some limitations are still present in the LQG description of BHs spacetimes. In relation to the SDBH model, an important limitation stays on the reduction of the microscopic degrees of freedom, due to the additional symmetries we have in LQC. In this context, one wonders if such a reduction could affect the final physical predictions of the theory when compared with a scenario constructed by taking into account the full LQG degrees of freedom.
A second important drawback in the SDBH scenario appears due to the fixing, by hand, of the parameters that are used in the construction of the states, which is performed at the end of the quantization procedure. It consists of an ambiguous prescription of the theory since different choices of the parameters can yield different physical scenarios.

In this way, a lot of effort to construct a more complete description of BHs, than that given by SDBHs, has been done in LQG, where more recent and improved approaches have been obtained. ( See,  e.g., \cite{Corichi:2015xia, Cortez:2017alh, Olmedo:2017lvt, Yonika:2017qgo, Ashtekar:2018lag, Ashtekar:2018cay, Alesci:2019pbs, Ashtekar:2020ckv} for more recent works). In this way, future investigations must address the issue of quasinormal modes and isospectrality breaking in the context of more recent and improved treatments to BHs in LQG.

{\acknowledgements}

We would like to thank CNPq, CAPES and PRONEX/CNPq/FAPESQ-PB (Grant no. 165/2018), for partial financial support. MBC and FAB acknowledge support from CNPq (Grant nos.  150479/2019-0, 312104/2018-9).


\begin{thebibliography}{99}

\bibitem{Cruz:2015bcj}
M.~B.~Cruz, C.~A.~S.~Silva and F.~A.~Brito,
Eur. Phys. J. C \textbf{79} (2019) no.2, 157
doi:10.1140/epjc/s10052-019-6565-2
[arXiv:1511.08263 [gr-qc]].

\bibitem{Einstein:1915ca}
A.~Einstein,
Sitzungsber. Preuss. Akad. Wiss. Berlin (Math. Phys. ) \textbf{1915} (1915), 844-847.

\bibitem{Abbott:2016blz}
B.~P.~Abbott \textit{et al.} [LIGO Scientific and Virgo],
Phys. Rev. Lett. \textbf{116} (2016) no.6, 061102
doi:10.1103/PhysRevLett.116.061102
[arXiv:1602.03837 [gr-qc]].

\bibitem{TheLIGOScientific:2016htt}
B.~P.~Abbott \textit{et al.} [LIGO Scientific and Virgo],
Astrophys. J. Lett. \textbf{818} (2016) no.2, L22
doi:10.3847/2041-8205/818/2/L22
[arXiv:1602.03846 [astro-ph.HE]].

\bibitem{Abbott:2016nmj}
B.~P.~Abbott \textit{et al.} [LIGO Scientific and Virgo],
Phys. Rev. Lett. \textbf{116} (2016) no.24, 241103
doi:10.1103/PhysRevLett.116.241103
[arXiv:1606.04855 [gr-qc]].

\bibitem{Abbott:2017vtc}
B.~P.~Abbott \textit{et al.} [LIGO Scientific and VIRGO],
Phys. Rev. Lett. \textbf{118} (2017) no.22, 221101
doi:10.1103/PhysRevLett.118.221101
[arXiv:1706.01812 [gr-qc]].

\bibitem{LIGOScientific:2020stg}
R.~Abbott \textit{et al.} [LIGO Scientific and Virgo],
[arXiv:2004.08342 [astro-ph.HE]].

\bibitem{Modesto:2009ve}
L.~Modesto and I.~Premont-Schwarz,
Phys. Rev. D \textbf{80} (2009), 064041
doi:10.1103/PhysRevD.80.064041
[arXiv:0905.3170 [hep-th]].

\bibitem{Hawking:1974sw}
S.~W.~Hawking,
Commun. Math. Phys. \textbf{43} (1975), 199-220
doi:10.1007/BF02345020.

\bibitem{Horowitz:1999jd}
G.~T.~Horowitz and V.~E.~Hubeny,
Phys. Rev. D \textbf{62} (2000), 024027
doi:10.1103/PhysRevD.62.024027
[arXiv:hep-th/9909056 [hep-th]].

\bibitem{Berti:2009kk}
E.~Berti, V.~Cardoso and A.~O.~Starinets,
Class. Quant. Grav. \textbf{26} (2009), 163001
doi:10.1088/0264-9381/26/16/163001
[arXiv:0905.2975 [gr-qc]].

\bibitem{Dreyer:2002vy}
O.~Dreyer,
Phys. Rev. Lett. \textbf{90} (2003), 081301
doi:10.1103/PhysRevLett.90.081301
[arXiv:gr-qc/0211076 [gr-qc]].

\bibitem{Santos:2015gja}
V.~Santos, R.~V.~Maluf and C.~A.~S.~Almeida,
Phys. Rev. D \textbf{93} (2016) no.8, 084047
doi:10.1103/PhysRevD.93.084047
[arXiv:1509.04306 [gr-qc]].

\bibitem{Cardoso:2019mqo}
V.~Cardoso, M.~Kimura, A.~Maselli, E.~Berti, C.~F.~B.~Macedo and R.~McManus,
Phys. Rev. D \textbf{99}, no.10, 104077 (2019)
doi:10.1103/PhysRevD.99.104077
[arXiv:1901.01265 [gr-qc]].

\bibitem{Moulin:2019bfh}
F.~Moulin and A.~Barrau,
[arXiv:1906.05633 [gr-qc]].



\bibitem{Modesto:2008im}
L.~Modesto,
Int. J. Theor. Phys. \textbf{49} (2010), 1649-1683
doi:10.1007/s10773-010-0346-x
[arXiv:0811.2196 [gr-qc]].

\bibitem{Zerilli:1970se}
F.~J.~Zerilli,
Phys. Rev. Lett. \textbf{24} (1970), 737-738
doi:10.1103/PhysRevLett.24.737.

\bibitem{Zerilli:1971wd}
F.~J.~Zerilli,
Phys. Rev. D \textbf{2} (1970), 2141-2160
doi:10.1103/PhysRevD.2.2141.

\bibitem{Schutz:1985km}
B.~F.~Schutz and C.~M.~Will,
Astrophys. J. Lett. \textbf{291} (1985), L33-L36
doi:10.1086/184453.

\bibitem{Iyer:1986np}
S.~Iyer and C.~M.~Will,
Phys. Rev. D \textbf{35} (1987), 3621
doi:10.1103/PhysRevD.35.3621.

\bibitem{Konoplya:2004ip}
R.~A.~Konoplya,
J. Phys. Stud. \textbf{8} (2004), 93-100.

\bibitem{Cardoso:2003cj}
V.~Cardoso, R.~Konoplya and J.~P.~S.~Lemos,
Phys. Rev. D \textbf{68} (2003), 044024
doi:10.1103/PhysRevD.68.044024
[arXiv:gr-qc/0305037 [gr-qc]].

\bibitem{Konoplya:2011qq}
R.~A.~Konoplya and A.~Zhidenko,
Rev. Mod. Phys. \textbf{83} (2011), 793-836
doi:10.1103/RevModPhys.83.793
[arXiv:1102.4014 [gr-qc]].

\bibitem{Kokkotas:1999bd}
K.~D.~Kokkotas and B.~G.~Schmidt,
Living Rev. Rel. \textbf{2} (1999), 2
doi:10.12942/lrr-1999-2
[arXiv:gr-qc/9909058 [gr-qc]].

\bibitem{Nollert:1999ji}
H.~P.~Nollert,
Class. Quant. Grav. \textbf{16} (1999), R159-R216
doi:10.1088/0264-9381/16/12/201.

\bibitem{Corichi:2015xia}
A.~Corichi and P.~Singh,
Class. Quant. Grav. \textbf{33}, no.5, 055006 (2016)
doi:10.1088/0264-9381/33/5/055006
[arXiv:1506.08015 [gr-qc]].

\bibitem{Cortez:2017alh}
J.~Cortez, W.~Cuervo, H.~A.~Morales-Técotl and J.~C.~Ruelas,
Phys. Rev. D \textbf{95}, no.6, 064041 (2017)
doi:10.1103/PhysRevD.95.064041
[arXiv:1704.03362 [gr-qc]].

\bibitem{Olmedo:2017lvt}
J.~Olmedo, S.~Saini and P.~Singh,
Class. Quant. Grav. \textbf{34}, no.22, 225011 (2017)
doi:10.1088/1361-6382/aa8da8
[arXiv:1707.07333 [gr-qc]].

\bibitem{Yonika:2017qgo}
A.~Yonika, G.~Khanna and P.~Singh,
Class. Quant. Grav. \textbf{35}, no.4, 045007 (2018)
doi:10.1088/1361-6382/aaa18d
[arXiv:1709.06331 [gr-qc]].

\bibitem{Ashtekar:2018lag}
A.~Ashtekar, J.~Olmedo and P.~Singh,
Phys. Rev. Lett. \textbf{121}, no.24, 241301 (2018)
doi:10.1103/PhysRevLett.121.241301
[arXiv:1806.00648 [gr-qc]].

\bibitem{Ashtekar:2018cay}
A.~Ashtekar, J.~Olmedo and P.~Singh,
Phys. Rev. D \textbf{98}, no.12, 126003 (2018)
doi:10.1103/PhysRevD.98.126003
[arXiv:1806.02406 [gr-qc]].

\bibitem{Alesci:2019pbs}
E.~Alesci, S.~Bahrami and D.~Pranzetti,
Phys. Lett. B \textbf{797}, 134908 (2019)
doi:10.1016/j.physletb.2019.134908
[arXiv:1904.12412 [gr-qc]].

\bibitem{Ashtekar:2020ckv}
A.~Ashtekar and J.~Olmedo,
[arXiv:2005.02309 [gr-qc]].

\end{thebibliography}
\end{document}